# Stellar Models with Microscopic Diffusion and Rotational Mixing II: Application to Open Clusters

Brian Chaboyer[1,2], P. Demarque[1,3] and M.H. Pinsonneault[1,4]

**ABSTRACT**

Stellar models with masses ranging from 0.5 to 1.3 $M_\odot$ were constructed in order to compare to young cluster observations of Li and of rotation velocities. The amount of Li depletion in cool stars is sensitive to the amount of overshoot at the base of the surface convection zone, and the exact metallicity of the models. Even when this uncertainty is taken into account, the Li observations are a severe constraint for the models and rule out standard models, and pure diffusion models. Stellar models which include diffusion and rotational mixing in the radiative regions of stars are able to simultaneously match the Li abundances observed in the Pleiades, UMaG, Hyades, Praesepe, NGC 752 and M67. They also match the observed rotation periods in the Hyades. However, these models are unable to simultaneously explain the presence of the rapidly rotating late G and K stars in the Pleiades and the absence of rapidly rotating late F and early G stars.

*Subject headings:* stars: interiors – stars: rotation – stars: abundances

[1] Department of Astronomy, and Center for Solar and Space Research, Yale University, Box 208101, New Haven, CT 06520-8101
[2] CITA, 60 St. George St., University of Toronto, Toronto, Ontario, Canada M5S 1A7
 Electronic Mail – I:chaboyer@cita.utoronto.ca
[3] Center for Theoretical Physics, Yale University
[4] present address: Department of Astronomy, Ohio State University, 174 W. 18th Ave., Columbus, OH 43210-1106

# 1. Introduction

Standard stellar evolution models have been very successful in explaining the morphology of the Hertzsprung-Russell diagram. However, as new types of observations have become available, it has become clear that the standard models are missing a key ingredient – the fact that stars transport material in stellar radiative regions. The evidence for particle transport in the radiative regions of stars comes from observations of $^7$Li in open cluster stars[1] (see Pinsonneault 1994). $^7$Li is an excellent tracer of particle transport in stellar radiative regions as it is easily destroyed at temperatures of $\sim 2.6 \times 10^6$ K. Thus, there exists a radius below which no $^7$Li exists. If this radius lies below the surface convection zone of a star then the $^7$Li observed at the surface of the star should not change with time. Recent observational studies have provided convincing evidence that main sequence $^7$Li depletion occurs in moderate mass ($M \lesssim 1.3\ M_\odot$) stars (Hobbs & Pilachowski 1988; Boesgaard 1991; Thorburn et al. 1993; Soderblom et al. 1993a). Significantly, beryllium is also observed to be depleted in the Sun (Anders and Grevesse 1989 and references therin) and in some field stars (Boesgaard 1976). Because beryllium burns at higher temperatures than $^7$Li, this indicates that particle transport takes place throughout the envelope (albeit at a reduced rate at increased depth).

Main sequence $^7$Li depletion is in clear contradiction with standard stellar evolution models. Several mechanisms have been advanced to explain this depletion: main sequence mass loss, microscopic diffusion of $^7$Li out of the surface convection zone, and mixing driven by rotation. Swenson & Faulkner (1992) found that $^7$Li depletion induced by main sequence mass loss has serious problems when applied to cluster observations. Thus, the $^7$Li observations provide compelling evidence that some form of particle transport occurs in the radiative regions of stars.

Evidence that rotation may be important in depleting $^7$Li comes from detailed studies of the Pleiades (Soderblom et al. 1993a), the Hyades (Thorburn et al. 1993) and the Praesepes (Soderblom et al. 1993b) which found that stars with the same mass, composition and age could have different $^7$Li abundances. Both standard stellar evolution models, and models which include diffusion predict that the $^7$Li depletion in a star is a unique function of age, mass and metallicity. Some other property must be important in determining the amount of particle transport within the radiative region of a star. A logical candidate for this property is the initial rotation velocity, for it is observed that T-Tauri stars with similar masses have a wide range in rotation velocities (Bouvier 1991). The fact that rotation can induce mass motions in stellar radiative regions has been known for some time (Eddington 1925; Sweet 1950; see Zahn 1993 for a recent review), however these motions have typically been ignored in stellar evolution calculations (but see Charbonneau & Michaud 1991; Charbonneau 1992).

Pinsonneault et al. (1989) and Pinsonneault, Kawaler & Demarque (1990) presented detailed stellar evolution models which included the effects of rotational mixing. Our philosophy is to treat the problem of stellar models with rotation as an initial value problem, in the spirit of Endal and Sofia (1976; 1978; 1981). We specify the initial conditions, and a set of physical assumptions about angular momentum loss, transport, and particle transport. We then calibrate our models on the Sun, requiring that the surface rotation rate and lithium depletion match the solar values. In practice, this determines the constant in the angular momentum loss law and the ratio between the diffusion coefficients for angular momentum transport and mixing. We then apply the same set of physics to models with a range in initial mass and angular momentum and explore the observational consequences of changes in the physical assumptions. For simplicity, we have assumed that parameters in our models are the same (and constant) regardless of the mass or age of the star being modeled. The situation inside an actual star may be considerably more complicated.

The surface rotation rates and light element abundances of low mass stars as a function of mass and age are sensitive to changes in the input physics and initial conditions. The observations allow us to rule out some classes of models, and can also be used as a guide to which uncertainties in the models have a significant impact on observable quantities. In previous work, we have studied rotational models of the Sun (Pinsonneault et al. 1989) and open cluster stars (Pinsonneault et al. 1990). We begin by summarizing previous results, and then discuss the consequences of including microscopic diffusion and our improvements in the input physics.

Angular momentum loss drives the rotational evolution of low mass stars. A general expression for the angular momentum loss rate is a constant multiplied by the surface angular velocity to some power.

---

[1] In stars, Li may exist in two different isotopes: $^6$Li and $^7$Li. It is extremely difficult observationally to distinguish between $^6$Li and $^7$Li, thus the observations typically determine the total Li present in a star. In meteorites, 7.5% of the Li atoms are $^6$Li (Anders & Grevesse 1989). Thus, it appears that the total Li content in a star is dominated by $^7$Li. We will assume that the Li observations actually measure $^7$Li.

The exponent affects the time dependence of the spin down at late ages (Kawaler 1988), while the constant is adjusted to reproduce the Sun. Efficient angular momentum loss in rapid rotators efficiently spun them down during the pre-MS, resulting in maximum ZAMS rotation rates of order 50 km/s at 30 Myr and 20 km/s at 70 Myr for solar mass models; this is in contradiction with observations in young stars (Stauffer and Hartmann 1987). A wide range in initial angular momentum was reduced to a small range in surface rotation in $\sim 2 \times 10^8$ yr, in agreement with rotation periods in the Hyades (Radick et al. 1987). These properties persisted for a wide range in parameter space.

Models with different initial angular momentum experience different degrees of rotational mixing; the time scale of angular momentum transport affects the timing of the mixing but not the overall amount in solar age systems because of the solar calibration. The timing of the lithium depletion was found to be consistent with the open cluster data in solar analogues, but the models of late F stars in the Hyades were over-depleted in Li with respect to the data - indicating potential problems with the mass dependence of the particle transport. The small data sets available at the time, did not allow for a quantitative comparison of the predicted and observed dispersion in lithium caused by mixing, although the existence of a dispersion could already be seen in some systems.

In these models which included rotational mixing, but not diffusion, the sensitivity of transport to gradients in the mean molecular weight were found to have only a small impact on the surface properties because in standard stellar models such gradients are only large in the deep core, which is far from the region where lithium is destroyed and has only a small moment of inertia. Since the work of Pinsonneault et al. (1989; 1990) was completed it has become clear that microscopic diffusion could have an important impact on stellar models (e.g. Proffitt & Michaud 1991). In addition, relatively easy to use formulae for the microscopic diffusion coefficients have become available (Bahcall & Loeb 1990; Michaud & Proffitt 1993). For this reason, a study of the combined effects of rotation and diffusion was undertaken. In addition, the opportunity was taken to update the physics (both rotational and standard) used within the code. In particular, a saturation level to the angular momentum loss law has been added, in order to allow for a wide range of rotation velocities near the ZAMS. These changes have had a significant impact on the models.

Microscopic diffusion leads to the separation of hydrogen and helium, naturally leading to sharp gradients in mean molecular weight at the base of the convection zone (see Vauclair 1983; Michaud and Charbonneau 1991 for good discussions of the underlying physics). Microscopic diffusion also causes $^7$Li to sink with respect to hydrogen in stellar envelopes (see Vauclair 1983, Michaud and Charbonneau 1991 for good discussions of the underlying physics) when it is fully ionized. For surface convection zones which are sufficiently thin, $^7$Li can be partially ionized and radiation pressure can cause $^7$Li to rise. Michaud (1986) proposed diffusion processes as the explanation for the strong depletion of $^7$Li seen in mid-F stars in the Hyades by Boesgaard and Trippico (1986). The timescale of diffusion increases with increased surface convection zone depth, making diffusion an unlikely explanation for later-type stars.

Chaboyer, Demarque & Pinsonneault (1994, Paper I) examined the effect of varying the sensitivity of rotational transport to gradients in the mean molecular weight. This degree of sensitivity was parameterized by $f_\mu$, large values of $f_\mu$ imply that the rotational mixing is strongly inhibited by gradients in mean molecular weight. In Paper I we found that $f_\mu \lesssim 0.1$ is required for appreciable main sequence $^7$Li depletion to occur. This is due to the fact that diffusion can lead to a sharp gradient in mean molecular weight at the base of the surface convection zone, which effectively halts the rotational mixing to the surface.

In this paper, we will examine the $^7$Li depletion due to diffusion and rotational mixing and compare to observations of $^7$Li abundances and surface rotation velocities in young cluster stars. Charbonneau & Michaud (1991) and Charbonneau (1992) studied the combined effect of diffusion and advection due to meridonal circulation (Charbonneau & Michaud 1991; Charbonneau 1992). Charbonneau & Michaud treated the meridonal circulation as a advection process; we treat meridonal circulation as a diffusion process. This is in agreement with the work of Chaboyer & Zahn (1992), who showed that appropriate rescaling of the turbulent diffusion coefficient allows the combined advection/diffusion process induced by meridonal circulation to be treated solely as a diffusion process. This work includes the effects of angular momentum loss, as well as many rotational instabilities which was not studied by Charbonneau & Michaud.

This paper is a follow up to Paper I which examined the interaction of diffusion and rotation in the Sun. Observations of young cluster stars provide an excellent test of stellar models. Stars in a given cluster have similar ages, distances and metallicities. These quantities are known far more precisely for cluster stars than for field stars. By studying a single cluster, it is possible to obtain information regarding the mass dependence of the $^7$Li depletion and surface rotation velocities. By comparing clusters of different ages, it is possible to determine how the $^7$Li depletion and surface rotation velocities vary with time. For this reason, a comparison between models and observations in young clusters provides a crucial diagnostic of particle transport which occurs in the radiative regions of stars. Our angular momentum loss law is inapplicable to early-type stars, and we do not include radiative levitation in our models; for these reasons, we concentrate on models with masses of 1.3 $M_\odot$ and lower. Hence, we will concentrate on comparing our models to the observations of $^7$Li abundances and rotation velocities in young stars with $M \leq 1.3$ $M_\odot$. A related paper, Chaboyer & Demarque (1994) examines the constraints that the observed halo star $^7$Li abundances put on particle transport in stellar radiative regions.

Although cluster observations contain a great deal of information regarding the timescale and mass dependence of $^7$Li depletion, the Sun provides a unique calibration of stellar models since its properties (age, chemical composition, luminosity) are known far more accurately than any other star. This allows one to calibrate the mixing length parameter which is used to model convection. By comparing photospheric and meteoritic abundances it is possible to determine the amount of destruction of $^7$Li. The Sun is the only star for which it is possible to make this measurement. For this reason, the Sun offers a unique opportunity to calibrate some of the parameters which are present in stellar models. Hence, all of the models we present in this paper will be calibrated to match the observed solar parameters, as discussed in Paper I. Full details of the stellar model construction, including descriptions of the particle and angular momentum transport and angular momentum loss law are found in Paper I. The mixing parameters and the properties of our calibrated solar models are shown in Table 1 of Paper I. Each set of mixing parameters was given a two letter code which will be used to refer to them. In these calibrated models, the mixing length, solar helium content and total metallicity have been varied in order to match the observed solar luminosity, radius and surface heavy element to hydrogen ratio. In addition, models which include rotation have been calibrated to yield the present solar surface rotation rate and $^7$Li depletion factor by varying the efficiency of mixing of chemical species, and an multiplicative constant in the angular momentum loss law.

In §2. we provide a brief description of the observational database which is used to test the models. The construction of stellar models and isochrones is discussed in §3. where some standard stellar models are presented and shown to disagree with the $^7$Li observations. Models which only include microscopic diffusion are presented in §4. In §5. the combined rotation and diffusion models are presented and found to be in reasonable agreement with the young cluster observations. Finally, the main results of this paper are summarized in §6.

## 2. The Observational Database

Observations of $^7$Li and surface rotation velocities in young clusters have the potential to delineate the time and mass dependence of angular momentum and mass transport within stars. Due to the advent of high efficiency CCD spectrographs, the database of cluster stars with observed $^7$Li abundances and rotational velocities has grown enormously in the last few years. The Pleiades, Hyades and Praesepe clusters have been extensively studied. The Pleiades has an age of 70 Myr which is particularly interesting as it implies that the solar mass stars have just arrived on the main sequence. Hence, observations in the Pleiades give us information on the pre-main sequence $^7$Li depletion and angular momentum loss. The Hyades and Praesepe are ten times as old, 700 Myr (Barry, Cromwell, & Hege 1987) and so yield information on processes which occur on the main sequence. One complicating factor is that the Pleiades has a metallicity of [Fe/H] = $-0.034 \pm 0.024$ (Boesgaard & Friel 1990), while the Hyades is more metal rich, [Fe/H] = $+0.127 \pm 0.022$ (Boesgaard & Friel 1990). Praesepe has a metallicity of [Fe/H] = $+0.038 \pm 0.039$ (Friel & Boesgaard 1992). In the Pleiades, high quality $^7$Li observations exist for 128 stars (Soderblom *et al.* 1993a). Accurate $v \sin i$ detections have been made for 89 stars and upper limits of $7 - 10$ km/s exist for 61 stars (Soderblom *et al.* 1993c; we have excluded stars with high reddennings). In the Hyades, a uniform set of high quality $^7$Li abundances for 52 stars and upper limits for 13 stars has been obtained by Thorburn *et al.* (1993). The Hyades are unique among the clusters in that rotational periods exist for 22 stars (Radick *et al.* 1987). The fact that we do not have to worry about inclination effects makes the use of rotation periods far superior to $v \sin i$ observations. Accurate $^7$Li abundances have been obtained for 63 stars in the Praesepe by Soderblom *et al.* (1993b). The Praesepe cluster has an age very similar to the Hyades, so it is interesting to inter-compare these two clusters.

In addition to the large data bases for the Pleiades, Hyades and Praesepe, there are a few other clusters which have interesting ages for which some observations are available. There are no nearby clusters with an age of 200 to 300 Myr; however we are in a midst of a group of stars whose common origin can be discerned from their common kinematics and chromospheric emission. The Ursa Major Group (UMaG) has an age of 300 Myr (Tomkin & Popper 1986, Soderblom & Mayor 1993) and a metallicity of [Fe/H] = $-0.08 \pm 0.09$ (Soderblom & Mayor 1993). Soderblom et al. (1993d) have obtained $^7$Li observations for 14 probable members of the UMaG which have high quality colors. There are few observations of $^7$Li in cluster stars which are older than the Hyades. The cluster NGC 752 has an age of 1.7 Gyr (Barry et al. 1987) and there are 12 $^7$Li detections and 5 upper limits (Hobbs & Pilachowski 1986a) for this cluster which has solar metallicity. Finally, M67 has an age which is roughly solar, $4.0^{+1}_{-0.5}$ Gyr (Demarque, Green & Guenther 1992) and a similar metallicity to the Sun, [Fe/H] = $-0.04\pm0.12$ (Hobbs & Thorburn 1991). There are 9 $^7$Li detections and 10 upper limits available for M67 (Hobbs & Pilachowski 1986b; Spite et al. 1987). We have elected not to use the data from the cluster $\alpha$ Per, due to difficulties with patchy reddenning and membership (Balachandran 1988).

## 3. Standard Models

In order to explore the effect of metallicity and overshoot on the models, a number of $^7$Li isochrones were constructed using standard physics (models GD, LA and PA from Table 1, Paper I). For each set of parameters we are interested in, a series of stellar models with masses 0.6, 0.7 ... 1.3 $M_\odot$ are evolved. The basic physics in these models are described in Paper I. In brief, the energy producing nuclear reaction rates are from Bahcall & Pinsonneault (1992); reaction rates for the $^6$Li, $^7$Li and $^9$Be are from Caughlan & Fowler (1988); the high temperature opacities from Iglesias & Rogers (1991); the low temperature opacities (below $10^4$ K) are from Kurucz (1991); while the surface boundary conditions are determined using Kurucz model atmospheres (Kurucz 1992; Howard, 1993). The fit between the stellar model and the atmosphere is made at an optical depth of $\tau = 2/3$. For temperatures above $10^6$ K, a relativistic degenerate, fully ionized equation of state is use. Below $10^6$ K, the single ionization of $^1$H, the first ionization of the metals and both ionizations of $^4$He are taken into account via the Saha equation.

Theoretical $^7$Li isochrones are constructed for a given age by linearly interpolating the $^7$Li depletion and $T_{\text{eff}}$ to the desired age for each individual mass. We shall refer to models with different masses but similar ages by the two letter code which is used in Table 1, Paper I to describe the parameters used in the construction of the solar models. Thus, when the term 'model JF isochrones' is used, we mean the isochrones which were constructed from models evolved using the parameters of the JF solar model.

In order to compare the $^7$Li destruction isochrones to the observations, it is necessary to assume an initial $^7$Li abundance. Observations of $^7$Li in T-Tauri stars are very difficult and produce a wide range of abundances, from $\log N(\text{Li}) = 2 \sim 4$, where $\log N(\text{Li}) \equiv \log(^7\text{Li}/\text{H})+12$ (Basri, Martin & Bertout 1991). The non-LTE work of Magazzu, Rebolo. and Pavlenko (1992) appears to have settled the question of the $^7$Li abundance in T-Tauri stars; these authors found $\log N(\text{Li}) = 3.2 \pm 0.3$ as the abundance of T-Tauri stars. The abundance of $^7$Li in the oldest meteorites is $\log N(\text{Li}) = 3.31 \pm 0.04$ (Anders & Grevesse 1988). Observations of $^7$Li in the interstellar medium yield $\log N(\text{Li}) = 3.53$ (Lemoine et al. 1993) with an likely error of $\pm 0.5$ dex (Hobbs, 1993) Thus, the initial $^7$Li abundance of our models is not a well determined quantity; it is likely to be in the range $\log N(\text{Li}) = 3.3 - 3.6$. We have chosen the initial $^7$Li abundance which yields the optimal fit to all of the cluster data and which lies within the range specified above. For models with non-solar metallicities, we have scaled the solar heavy element mixture using [Fe/H] = $\log(Z/Z_\odot)$ and calculated the $^4$He abundances by assuming $\Delta Y = 2.5 \Delta Z$.

Figure 1 demonstrates the sensitivity of the models to the amount of overshoot below the convection zone. Changing the overshoot by 0.05 pressure scale height ($H_p$) has a noticeable effect on the cool models. Models with no overshoot, or an overshoot of 0.02 pressure scale height, provide a good match to the mean trend of the Pleiades observations. However, the models do not account for the large spread in $^7$Li abundances which is observed below 5300 K. The standard models do **not** match the mean trend observed in the Hyades below 6000 K. The models are under-depleted with respect to the majority of the observations. This is a clear indication that mixing is active in stars between 70 and 700 Myr. *Note that the standard models* **do** *match the observed $^7$Li abundances in the tidally locked binaries. As rotational mixing is unlikely to occur in tidally locked binaries, this is evidence that the*

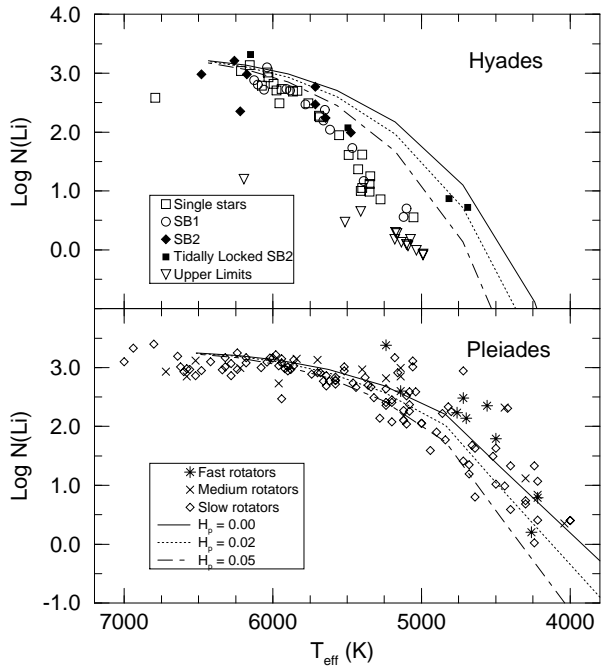

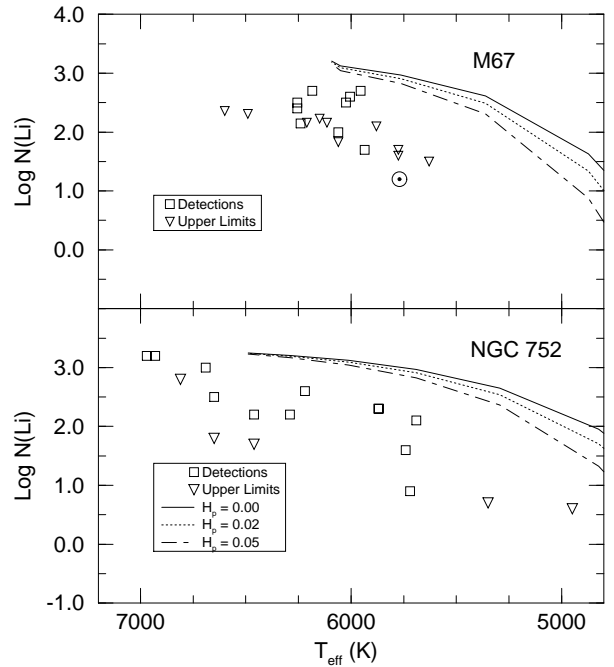

Fig. 1.— The effect of overshoot on the $^7$Li destruction in standard models is demonstrated by comparing $^7$Li destruction isochrones from models PA ($H_p$ = 0.00), LA ($H_p$ = 0.02) and GD ($H_p$ = 0.05) to the observations in the Pleiades and Hyades. The initial $^7$Li abundance has been taken to be $\log N(Li) = 3.3$. The observations are shown as various symbols, while the results of the model calculations are given as lines. The Pleiades models have [Fe/H] = 0.0, while the Hyades models have [Fe/H] = +0.1.

Fig. 2.— Standard models with different amounts of overshoot are compared to Li observations in M67 and NGC 752. The initial $^7$Li abundance has been taken to be $\log N(Li) = 3.3$. The observations are shown as squares and triangles, and the theoretical isochrones are given as lines. The $^7$Li abundance of the Sun is shown by the solar ($\odot$) symbol in the M67 graph. The NGC 752 isochrones have an age of 1.7 Gyr, while the M67 isochrones have an age of 4.0 Gyr.

*additional mixing mechanism which operates between the ages of the Pleiades and the Hyades is related to rotation.* The mismatch between standard models and observations becomes larger as we go to older stars. This is shown in Figure 2 where the theoretical isochrones are compared to observations in M67 and NGC 752. It is clear that standard models do not deplete enough $^7$Li to account for the observations. We note that a recent paper by Swenson *et al.* (1994) was able to match the observed $^7$Li abundances in single Hyades stars using standard models with enhanced oxygen abundances. However, as noted by Swenson *et al.* (1994), all outlying points represent spectroscopic or tidally locked binaries.

When comparing the theoretical isochrones to the observations, it is important to remember that the metallicities of the clusters are somewhat uncertain. Figure 3 demonstrates the dramatic effect changing the metallicity has on the theoretical isochrones. It is clear that a change in metallicity of 0.05 dex (which is within the $2\sigma$ errors of the observations) can change considerably whether or not we judge a model to provide a good fit to the data.

It is clear from Figures 1 – 3 that the amount of $^7$Li destruction in stellar evolutionary models is sensitive to the amount of overshoot included in the models and the exact metallicity of the models. Changes in metallicity of the magnitude we have studied here ($\Delta$[Fe/H] = 0.25) produce dramatic effects for stars with effective temperatures below 6000 K. Changing the extant of the overshoot layer by 0.05 pressure scale-height leads to noticeable effects for models with $T_{eff} \lesssim 5500$. Thus, it is the cool stars which are most affected by the uncertainties in the observed metallicity and extent of the overshoot layer. Changes in [Fe/H] which are within the $2\sigma$ errors of the observations can turn an unacceptably bad fit into a good fit. In addition to the problems with the theoretical isochrones, the observations have random and possible systematic errors associated with them. Due to this extreme sensitivity, we have decided not to perform a least squares fit of the $^7$Li destruction isochrones to the data. To do such an analysis would require us to put error bars on the $^7$Li destruction analysis, an endeavor which would require explor-ing an enormously large parameter space. Instead of a least squares analysis, we have merely chosen an initial $^7$Li abundance (in the range $\log N(Li) = 3.3 - 3.6$) which yields the optimal 'eye' fit to the Pleiades and Hyades $^7$Li data. When examining the $^7$Li destruction isochrones it is important to recall the uncertainties in the depletion of $^7$Li in the cool stars. It is clear however, that no change in metallicity or overshoot could make the standard models simultaneously fit the observed $^7$Li abundances in the Pleiades, Hyades, NGC 752 and M67.

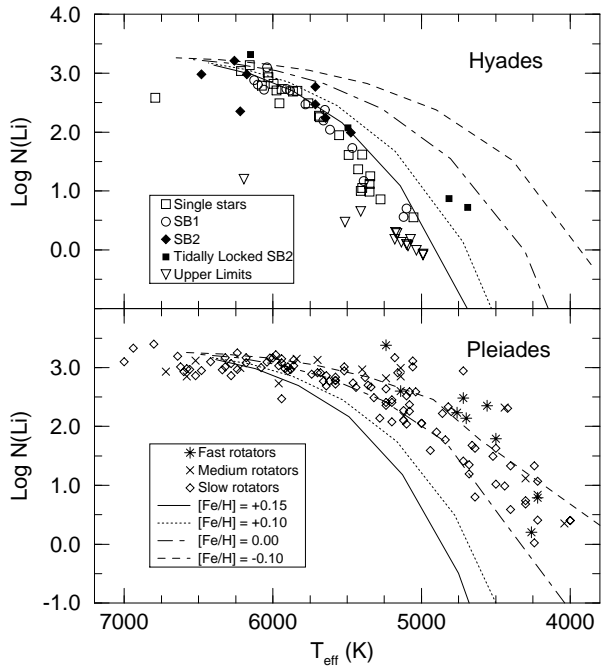

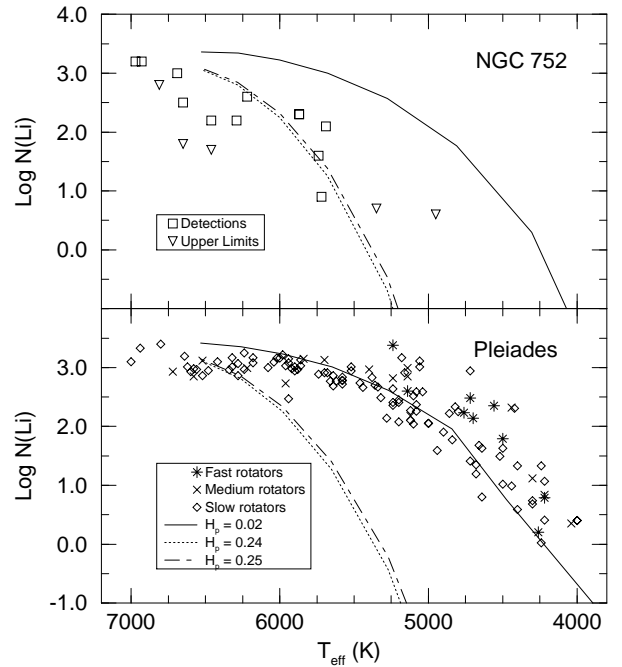

Fig. 3.— Standard models with varying metallicities and an overshoot of 0.05 pressure scale height (model GD) are compared to Li observations in the Hyades and the Pleiades. The initial $^7$Li abundance has been taken to be $\log N(Li) = 3.3$. The observations are shown as various symbols, and the theoretical isochrones are given as lines. The Pleiades has $[Fe/H] = -0.034 \pm 0.024$ while the Hyades has $[Fe/H] = 0.127 \pm 0.022$.

Fig. 4.— Models which include microscopic diffusion but do **not** include rotational mixing, are compared to Li observations in the Hyades and the NGC 752. The initial $^7$Li abundance has been taken to be $\log N(Li) = 3.5$. The observations are shown as various symbols, and the theoretical isochrones are given as lines. The isochrones shown are from models GD ($H_p = 0.02$), XB ($H_p = 0.24$) and XC ($H_p = 0.24$).

## 4. Pure Microscopic Diffusion Models

Models which include microscopic diffusion differ from standard models in that the diffusion of $^7$Li out of the surface convection zone leads to a depletion of $^7$Li at the surface on the main sequence. The models use the microscopic diffusion coefficients from Michaud & Proffitt (1993). The coefficients are uncertain at the $\sim 15\%$, for this reason, an adjustable constant $f_m$ has been introduced which is set to 0.8 or 1.0. In Figure 4, $^7$Li observations in the Pleiades and NGC 752 are compared to models which include microscopic diffusion but **not** rotational mixing. The model with an overshoot of 0.02 pressure scale height (model ND) provides a reasonable fit to the mean trend in the Pleiades, but does **not** match the observations in NGC 752. The $^7$Li depletion in models which include microscopic diffusion but not rotational mixing is very similar to the standard models. This is due to the long time scale of microscopic diffusion in the low mass stars which are studied here. Thus, these models have the same problem as the standard models do – there is not enough $^7$Li depletion on the main sequence.

Two models with large amounts of overshoot (0.24 and 0.25 pressure scale height) are also shown in Figure 4. The amount of overshoot in these models provides a good fit to the observed $^7$Li depletion in the Sun (models XB and XC). It is clear that such models do no fit the observations in the Pleiades; there has been far too much pre-main sequence depletion. The match to the observations in NGC 752 is reasonable. This is not too surprising, as we have calibrated the models to match the Sun at an age of 4.55

Gyr. Large amounts of overshoot has been suggested by Ahrens, Stix & Thorn (1992) as an explanation for the $^7$Li depletion in the Sun. This suggestion is only viable if we assume that the Pleiades (and other young clusters, such as $\alpha$ Per, whose stars has similar $^7$Li abundances to the Pleiades stars) have anomalously small amounts of overshoot compared to the older clusters NGC 752 and M67 and the Sun. This is clearly an unacceptable explanation, so we can rule out large amounts of overshoot as a reasonable explanation for the solar $^7$Li depletion.

## 5. Models with Microscopic Diffusion and Rotation

### 5.1. Parameter Variations

In this section, the effects of microscopic diffusion and rotational mixing are studied by comparing models with different parameters to observations of the Pleiades and Hyades stars. In order to make the inter-comparisons between different models straightforward, we will present isochrones that have the same metallicity and age. For the Pleiades, all isochrones have an age of 70 Myr and a solar metallicity while the Hyades isochrones have an age of 700 Myr and [Fe/H] = +0.10.

The details of the rotational mixing prescription are found in Paper I, here we provide a brief summary. The rotational mixing coefficients are taken to be the product of a velocity estimate and the radius (Zahn 1993). The velocity estimates are similar to those used by Pinsonneault et al. (1989) and include meridonal circulation, dynamical shear instabilities, the Solberg-Hoiland instability, the Goldreich-Schubert-Fricke instability and the secular shear instability. The rotating models include a general parameterization for the loss of angular momentum at the surface due to magnetic stellar winds. The formulation is similar to that given by Kawaler (1988), except that a saturation level, $\omega_{\rm crit}$ has been introduced. If the surface angular velocity is below $\omega_{\rm crit}$ then the rate of angular momentum loss is that given by Kawaler (1988). Above $\omega_{\rm crit}$, angular momentum loss occurs at a reduced level compared to that given by Kawaler (1988). In addition to $\omega_{\rm crit}$, there are four other free parameters in the rotating models: the initial rotation velocity given to the calibrated solar model ($V_i$); the power law index for in the angular momentum loss law ($N$); a constant which multiplies the diffusion coefficient for the Goldreich-Schubert-Fricke instability ($f_{GSF}$), and $f_\mu$ which characterizes the sensitivity of the rotational mixing to gradient in the mean molecular weight.

In order to construct the $^7$Li depletion and rotation velocity isochrones, a series of stellar models (masses $0.60, 0.70, \ldots 1.30$ $M_\odot$) were calculated with [Fe/H] = 0.0 and +0.10. For each mass and metallicity, 3 models were constructed, with initial rotation velocities of 10, 30 and 50 km/s. This spans the range observed in T-Tauri stars (Bouvier 1991, Bouvier et al. 1993), with the majority of T-Tauri stars having rotation velocities below 20 km/s. For each initial rotation velocity, the $^7$Li destruction and rotation rate isochrones were produced in the same manner as was done for the standard models. The Radick et al. (1987) rotation period data for the Hyades includes B-V colors, not effective temperatures. In order to compare the models to the observations of rotation rates, the effective temperatures have been converted to B-V colors using an updated version of the RYI color calibration (Green, Demarque & King 1987; Green 1988).

Surface rotation and $^7$Li depletion isochrones have been calculated for nearly all of the models presented in Table 1, Paper I (Chaboyer 1993). Here, only a selected subset of the models will be presented. We have selected some models which illustrate the sensitivity of the models to the various parameters, and present a model which best represents the observed $^7$Li depletion and rotation velocity patterns observed in young cluster stars. We first consider an $f_\mu = 0.10$ model, UU. These models have an overshoot of 0.05 pressure scale height, $N = 2$, $\omega_{\rm crit} = 3.0 \times 10^5$ $s^{-1}$ and the microscopic diffusion coefficients are given by Michaud & Proffitt (1993). The models are compared to observations in Figures 5 – 6. It is clear from these figures that larger initial rotation velocities lead to larger $^7$Li destruction. Figure 5 shows that in model UU, the cool Pleiades stars are over-depleted in $^7$Li, while the hot stars are over-depleted in the Hyades. The surface rotation isochrones are shown in Figure 6 with the Pleiades and the Hyades observations. When examining Figure 6 (and all subsequent rotation rate plots) it is important to realize that the Pleiades observations are $v \sin i$ data. No attempt has been made to correct for inclination effects. The fastest rotators have likely been observed with $\sin i = 1$. The average statistical factor which converts $v \sin i$ data to $v$ is $4/\pi$. The comparison to the rotation periods in the Hyades is relatively straightforward, as there are no inclination effects to worry about. Model UU fails to match the rapidly rotating cool stars in the Pleiades and the hot stars in the Hyades. The failure of model UU isochrones to fit the observed $^7$Li abundance in the hot stars in the Hyades and to match the rotation velocities of the cool stars in the Pleiades leads us to reject the UU models as viable stellar models. The other models with $f_\mu = 0.1$ presented in Table 1, Paper I did an equally poor (or worse) job of matching the observations.

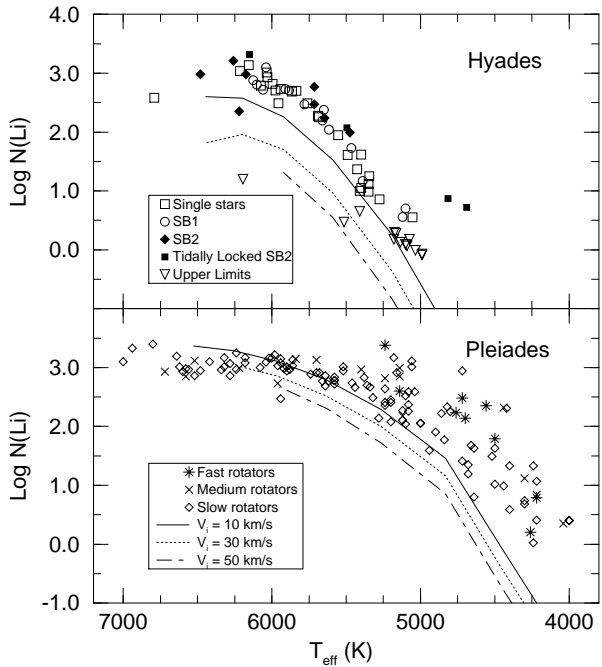
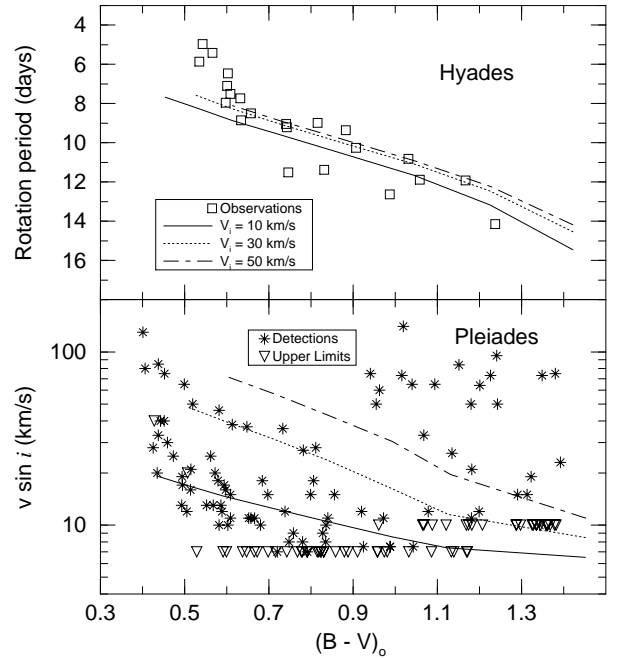

Fig. 5.— Models UU $^7$Li destruction isochrones are compared to $^7$Li observations in the Hyades and the Pleiades. The initial $^7$Li abundance has been taken to be $\log N(Li) = 3.5$. The observations are shown as various symbols. The theoretical isochrones are given as lines. The legend identifies the initial rotation velocity given to the models. The Pleiades models have a solar metallicity, while the Hyades models have [Fe/H] = +0.10.

Fig. 6.— Model UU surface rotation isochrones are compared to $v \sin i$ observations in the Pleiades and rotation period observations in the Hyades. The observations are shown as various symbols. The theoretical isochrones are given as lines and are identified by the initial rotation velocity given to the models.

A number of calibrated models have been calculated which have with $f_\mu = 0.01$, overshoot of 0.02 pressure scale height, and microscopic diffusion coefficients multiplied by 0.8. The match between the most promising models and the Hyades and Pleiades observations are shown in figures 7 – 11. The JF $^7$Li isochrones (with $f_{\rm GSF} = 10$, $\omega_{\rm crit} = 3.0 \times 10^{-5}\ s^{-1}$) shown in Figure 7 are less depleted than model UU. However, the JF isochrones still do not match the hot stars in the Hyades; they are over-depleted. JF rotational velocity isochrones are similar to the UU cases already shown and yield a poor fit to the observations.

The VN isochrones shown in Figures 8 and 9 have $f_{\rm GSF} = 1.0$, and $\omega_{\rm crit} = 1.5 \times 10^{-5}\ s^{-1}$ provide a better fit to the data than the previous cases we have shown. It is clear that changing the efficiency of rotational mixing and angular momentum loss has an important effect on the models. The VN $^7$Li isochrones show very little dispersion at the age of the Pleiades, in contrast to the observations. In addition, the fastest rotating models deplete the most Li, in contrast to the observations. However, the match to the Hyades is quite good. In particular, the depletion of $^7$Li in hot stars ($T_{\rm eff} \gtrsim 5500$) matches the observed $^7$Li abundances. The surface rotation isochrones are shown in Figure 9. Although there are still a few very rapidly rotating late type stars in the Pleiades which do not fit the models, the overall match to the observations is better than the previous cases. The predicted Hyades rotational periods match the observations well for stars with $B - V \gtrsim 0.6$. Stars hotter than this still rotate too fast with respect to the models. Overall, it is clear the the VN isochrones represent the best match to observations of the models considered thus far. Model WE has identical parameters to model VN, except it was assumed that the solar $^7$Li depletion was 100 for model WE (it is a factor of 200 in model VN). Although they are not shown here, the model WE $^7$Li isochrones are quite similar model VN. This indicates that our models are not too sensitive to the exact $^7$Li depletion assumed for the Sun.

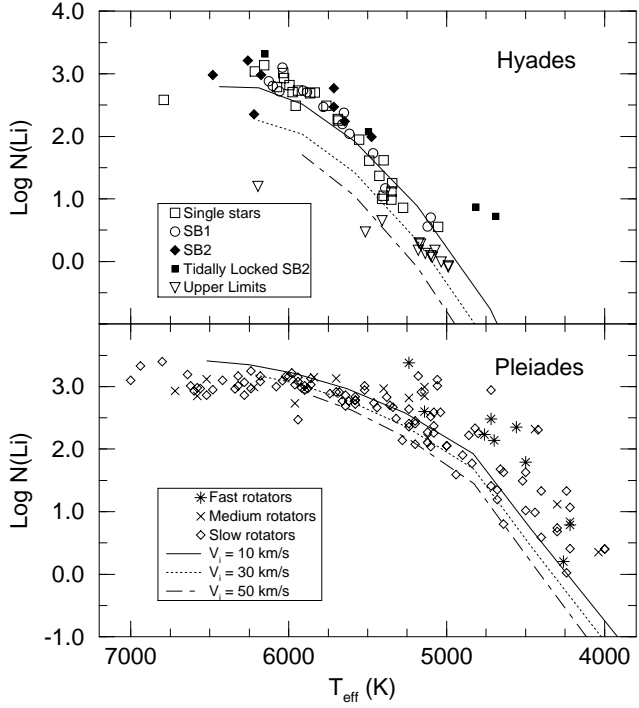 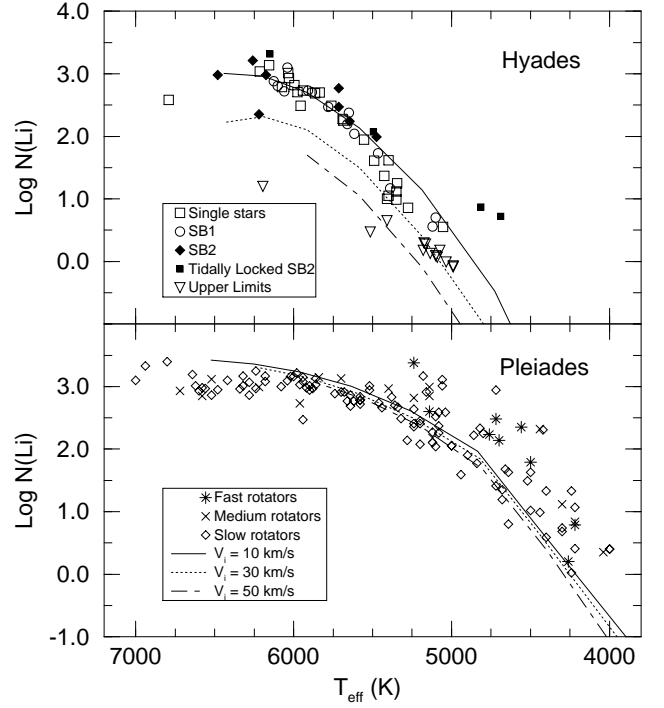

Fig. 7.— Model JF $^7$Li destruction isochrones are compared to $^7$Li observations in the Hyades and the Pleiades. The initial $^7$Li abundance has been taken to be $\log N(\text{Li}) = 3.5$. The observations are shown as various symbols. The theoretical isochrones are given as lines. The legend identifies the initial rotation velocity given to the models. The Pleiades models have a solar metallicity, while the Hyades models have [Fe/H] = +0.10.

Fig. 8.— Model VN $^7$Li destruction isochrones are compared to $^7$Li observations in the Hyades and the Pleiades. The initial $^7$Li abundance has been taken to be $\log N(\text{Li}) = 3.5$. The observations are shown as various symbols. The theoretical isochrones are given as lines. The legend identifies the initial rotation velocity given to the models. The Pleiades models have a solar metallicity, while the Hyades models have [Fe/H] = +0.10.

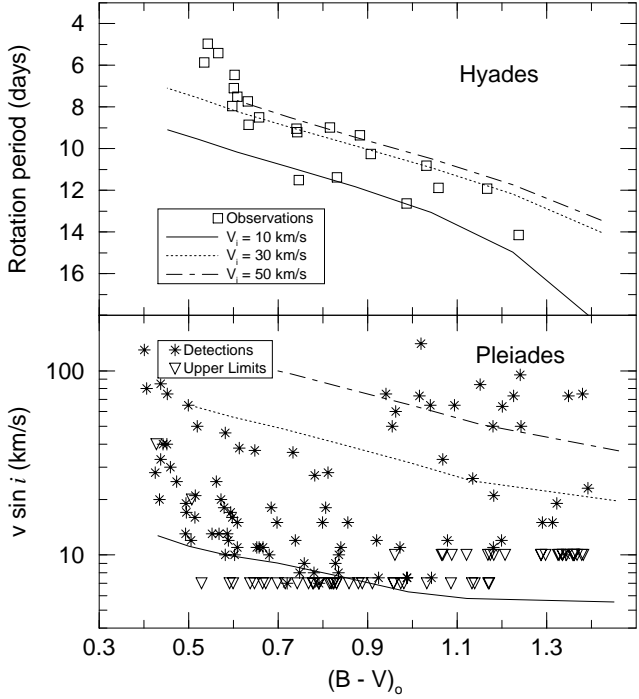
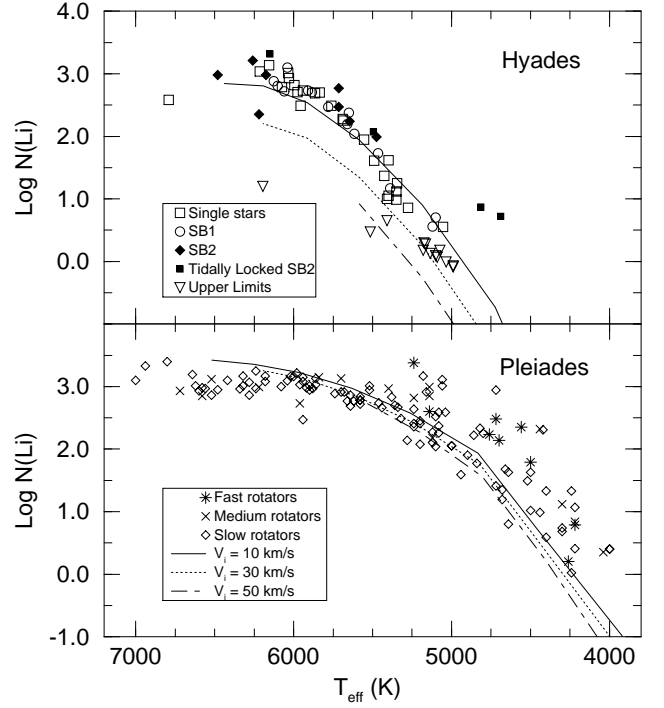

Fig. 9.— Model VN surface rotation isochrones are compared to $v \sin i$ observations in the Pleiades and rotation period observations in the Hyades. The observations are shown as various symbols. The theoretical isochrones are given as lines and are identified by the initial rotation velocity given to the models.

Fig. 10.— Model TJ $^7$Li destruction isochrones are compared to $^7$Li observations in the Hyades and the Pleiades. The initial $^7$Li abundance has been taken to be $\log N(Li) = 3.5$. The observations are shown as various symbols. The theoretical isochrones are given as lines. The legend identifies the initial rotation velocity given to the models. The Pleiades models have a solar metallicity, while the Hyades models have [Fe/H] = +0.10.

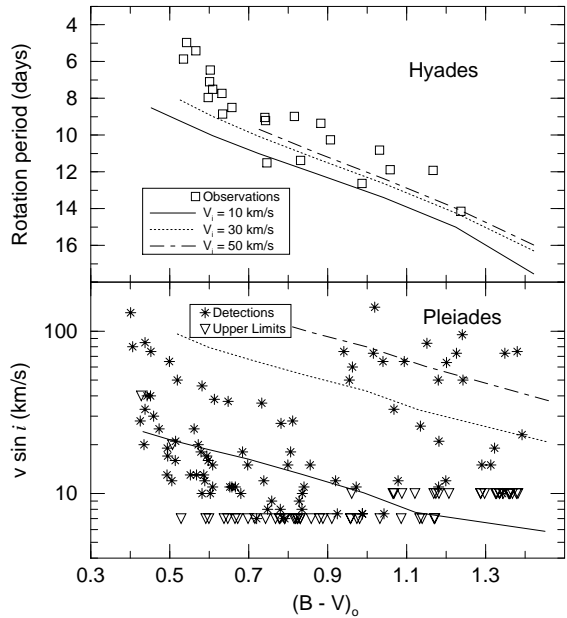

Fig. 11.— Model TJ surface rotation isochrones are compared to $v \sin i$ observations in the Pleiades and rotation period observations in the Hyades. The observations are shown as various symbols. The theoretical isochrones are given as lines and are identified by the initial rotation velocity given to the models.

Finally, the $^7$Li depletion model TJ isochrones are compared to observations in the Pleiades and the Hyades in Figure 10. Model TJ has $N = 1.8$ and $f_{GSF} = 10$, otherwise it is identical to model VN. It is clear that this parameter set (as do most of the previous ones) depletes too much $^7$Li in hot stars compared to the Hyades observations. In addition the rotation periods are too long compared to the Hyades observations (Figure 11). Hence, model TJ is not compatible with the observations. This demonstrates the importance the GSF diffusion coefficients have in determining the $^7$Li depletion and surface rotation rates of the models.

### 5.2. Optimal match to the Observations

In the previous section, the age and metallicity of the isochrones remained fixed as we compared various parameter sets to the observations in the Pleiades and the Hyades. It is clear from that study that the model VN isochrones were the best fit to the observations. An optimal match to the observations in the Pleiades, Hyades, Praesepe, UMaG, NGC 752 and M67 is made by varying the age and metallicity within the $1\sigma$ errors. A comparison to $^7$Li observations in the Praesepe and Hyades is shown in Figure 12. Figure

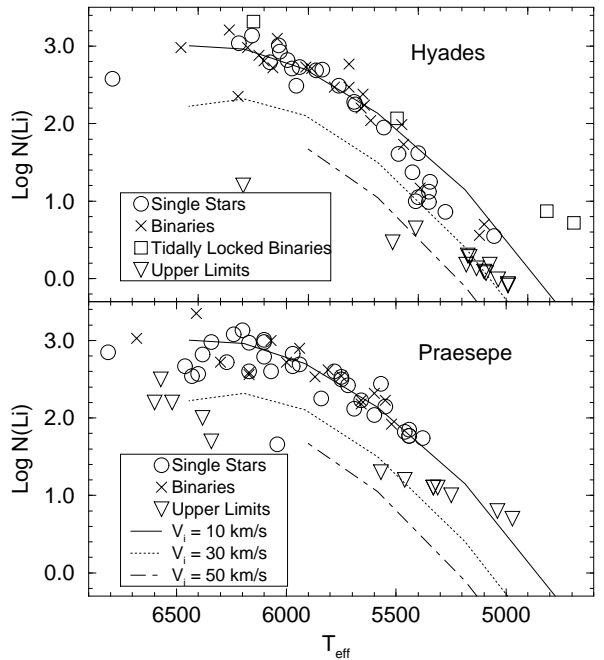

Fig. 12.— Model VN $^7$Li destruction isochrones are compared to $^7$Li observations in the Hyades and the Praespes. The initial $^7$Li abundance has been taken to be $\log N(Li) = 3.5$. The observations are shown as various symbols.

13 shows the fit to $^7$Li observations in the Pleiades, UMaG, NGC 752 and M67. From these two figures, it is clear that our models are capable of reproducing in detail the time and mass dependence of the observed $^7$Li abundances. However, the fit to the Pleiades does reveal a few problems. The dispersion (at a given $T_{eff}$) in the $^7$Li abundance is much smaller in the models than in the observations. In addition, there are a few stars (near 5000 K) which lie well above the isochrones, indicating that some mechanism is stopping the $^7$Li depletion in these stars which is not present in the models. The rotational velocity isochrones for the best fit cases are very similar to those already shown for model VN (Figure 9) and are not repeated here.

### 6. Summary

The observed lithium depletion pattern in low mass stars is strong evidence for physical processes not included in the standard stellar model. In this paper, we have considered the hypothesis that mixing induced by rotational instabilities and microscopic diffusion is responsible for the observed lithium depletion pattern. The empirical data requires a mixing mecha-nism which allows for the existence of a dispersion in abundance at fixed $T_{\text{eff}}$ in clusters and depletion at a declining rate with increased MS age. These are natural features of rotational mixing. The observed surface rotation rates can be used to constrain the angular momentum evolution of low mass stars, especially angular momentum loss rates. If the rotational mixing paradigm is accepted, then the surface lithium abundances can be used to constrain the time scales for mixing and internal angular momentum transport.

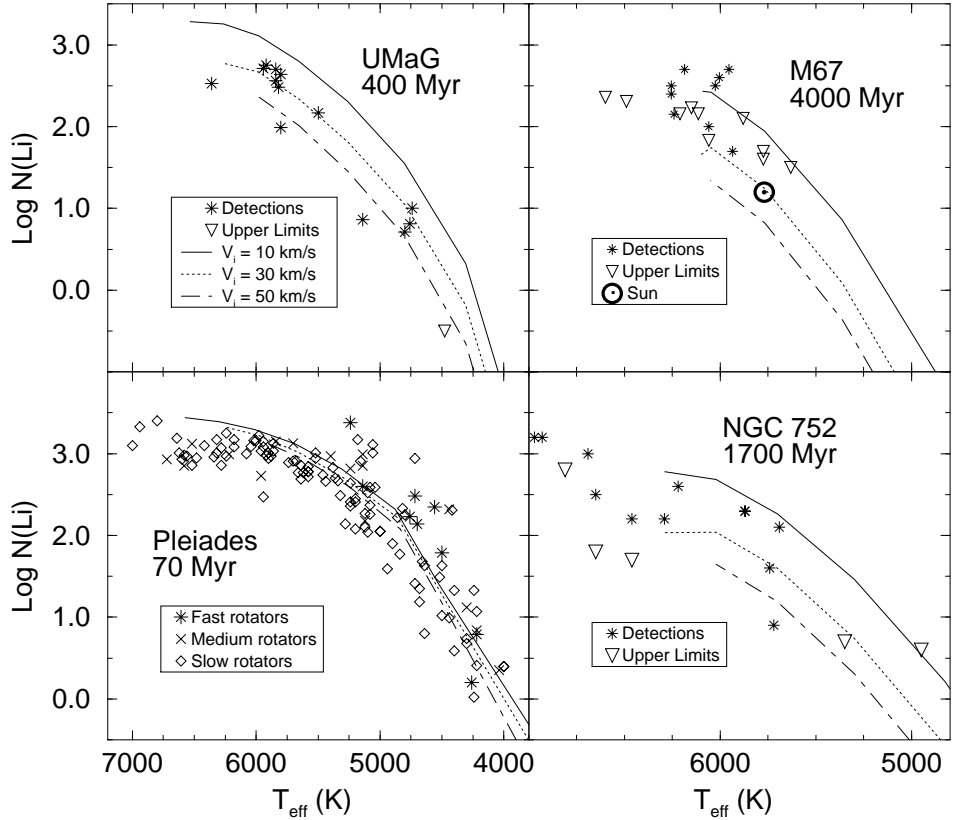

Fig. 13.— Model VN $^7$Li destruction isochrones are compared to $^7$Li observations in the Pleiades, UMaG, NGC 752 and M67. The initial $^7$Li abundance has been taken to be $\log N(\text{Li}) = 3.5$. The observations are shown as various symbols. The theoretical isochrones are given as lines. The ages and metallicities have been varied to yield an optimal fit to the data. The isochrone ages are given on the plot. The metallicities used are [Fe/H] $= -0.05$ for the Pleiades, and [Fe/H] $= 0.00$ the others.

The available data on rotation velocities and $^7$Li abundances in young clusters has been examined in order to constrain the various free parameters in our models. Standard models do not match the $^7$Li abundances observed in older clusters such as NGC 752 (age 1700 Myr) and M67 (age 4000 Myr). However, the standard models do match the mean trend in the $^7$Li abundances observed in the young clusters such as the Pleiades (age 70 Myr). This is clear evidence that some form of mixing occurs in the radiative regions of low mass main sequence stars. The addition of microscopic diffusion does not solve this problem. In the low mass stars studied here ($M \leq 1.3\ M_\odot$), the time scale for microscopic diffusion of $^7$Li is extremely long. Hence, the $^7$Li depletion in models which include microscopic diffusion are very similar to standard models.

We have constructed a large number of stellar models which include microscopic diffusion and rotationally induced mixing. Due to the uncertainties in the time scale for mixing of these instabilities and our primitive knowledge of angular momentum loss from the surface of a star, there are a number of free parameters associated with the models. When rotational mixing is included in the models, the match to the $^7$Li abundances in older clusters and the Sun becomes quite good. By fine tuning the parameters, it is possible to obtain a good fit to the mean $^7$Li abundances in clusters spanning a wide range in age.

The $^7$Li abundance measurements and surface rotation observations available for the Pleiades and the Hyades put tight constraints on the models. These observations rule out most of the parameter combinations we have attempted. The best match to the observations is obtained using model VN and is shown in Figures 9, 13 and 12. The models do a good job of matching the mean trend in the observed $^7$Li abundances in clusters with ages varying from 70 Myr to 4000 Myr. However, there are still some problems associated with the Pleiades models. In particular, there are a number of stars (with $T_{eff} \leq 5300$ K, $M \lesssim 0.9\ M_\odot$, many of which are fast rotators) which have higher $^7$Li abundances than is allowed by the models. It is clear that we are not modelling some mechanism which is halting the $^7$Li depletion in certain stars.

The VN isochrones are a good fit to the observed rotation periods in the Hyades. The observed $v \sin i$ rotation velocities in the Pleiades include a number of rapidly rotating cool stars ($B - V > 0.9$, $M \lesssim 0.9\ M_\odot$) which our models are unable to reproduce. The fact that the stars which are under-depleted in $^7$Li are in the same mass range is intriguing. It would appear that certain low mass stars have not suffered much angular momentum loss during the pre-main sequence phase. This could lead to a smaller amount of rotationally induced mixing within these stars, which would explain the population of stars which have higher than normal $^7$Li abundances at the age of the Pleiades. There are no such outliers observed in the Hyades, and so these stars must spin down between 70 and 700 Myr.

B.C. would like to thank Sabatino Sofia and Douglas Duncan, who provided him with their thoughtful comments on his thesis, which this paper is drawn from. In addition, B.C. is grateful to Charles Proffitt who provided him with detailed information regarding the microscopic diffusion coefficients used in this work. The comments of the anonymous referee improved the presentation of this paper. Research supported in part by NASA grants NAG5–1486, NAGW-2136, NAGW–2469 and NAGW-2531.

**REFERENCES**

Ahrens, B., Stix, M., & Thorn, M. 1992, A&A, 262, 673

Anders, E., & Grevesse, N. 1989, Geochim. Cosmochim. Acta, 53, 197

Bahcall, J.N., & Loeb, A. 1990, ApJ, 360, 267

Bahcall, J.N., & Pinsonneault, M.H. 1992, Reviews of Modern Physics, 64, 885

Balachandran, S. 1988, PhD thesis, UT Austin

Barry, D.C., Cromwell, R.H. & Hege, K. 1987, ApJ, 315, 264

Basri, G., Martin, E.L. & Bertout, C. 1991, A&A, 252, 625

Boesgaard, A. M. 1991, ApJL, 370, L95

Boesgaard, A. M. 1989, ApJ, 336, 798

Boesgaard, A.M. 1976, ApJ, 210, 466

Boesgaard, A. M., & Friel, E.D. 1990, ApJ, 351, 467

Bouvier, J. 1991, in Angular Momentum Evolution of Young Stars, ed. S. Catalano & J.R. Stauffer, (Dordrecht: Kluwer), 41

Bouvier, J., Cabrit, S., Fernández, Martín, E.L. and Matthews, J.M. 1993, A&A, 272, 176

Caughlan, G.R. & Fowler, W.A. 1988, Atomic Data & Nuclear Data Tables, 40, 283

Chaboyer, B.C. 1993, PhD thesis, Yale University

Chaboyer, B.C., Deliyannis, C. P., Demarque, P., Pinsonneault, M. H., & Sarajedini, A. 1992, ApJ, 388, 372

Chaboyer, B.C., Demarque, P. & Pinsonneault, M.H. 1994, submitted to ApJ (Paper I)

Chaboyer, B.C. & Zahn, J.-P. 1992, A&A, 253, 173

Charbonneau, P. 1992, A&A, 259, 134

Charbonneau, P. & Michaud, G. 1991, ApJ, 370, 693

Charbonneau, P. & MacGregor, K.B. 1993, ApJ, 417, 762

Demarque, P., Green, E.M. & Guenther, D.B. 1992, AJ, 103, 151

Eddington, A.S. 1925, Observatory, 48, 73

Endal, A.S. & Sofia, S. 1976, ApJ, 210, 184

Endal, A.S. & Sofia, S. 1978, ApJ, 220, 279

Endal, A.S. & Sofia, S. 1981, ApJ, 243, 625

Fricke, K. 1968, Zs Ap., 68, 317

Friel, E.D. & Boesgaard, A.M. 1992, ApJ, 387, 170

Goldreich, P. & Schubert, G. 1967, ApJ, 150, 571

Green, E.M. 1988, in Calibration of Stellar Ages, ed. A.G.D. Phillip (Schenectady, N.Y.:L. Davis Press) 81

Green, E.M., Demarque, P., & King, C.R. 1987, The Revised Yale Isochrones and Luminosity Functions (Yale Univ. Obs., New Haven)

Hobbs, L.M. 1993, private communication

Hobbs, L.M. and Pilachowski, C. 1988, ApJ, 334, 734

Hobbs, L. M. & Pilachowski, C. 1986a, ApJL, 309, L17

Hobbs, L. M. & Pilachowski, C. 1986b, ApJL, 311, L37

Hobbs, L. M. & Thorburn, J.A. 1991, AJ, 102, 1070

Howard, J.M. 1993, PhD thesis, Yale University

Iglesias, C.A. & Rogers, F.J. 1991, ApJ, 371, 408

Kawaler, S.D. 1988, ApJ, 333, 236

Kippenhahn, R. 1974, in IAU Symp. 66, Late Stages

of Stellar Evolution, ed. R. Tayler (Dordrecht: Reidel), 20

Kippenhahn, R., & Möllenhoff, C. 1974, Ap. Space Sci 31, 117

Kurucz, R.L. 1991, in Stellar Atmospheres: Beyond Classical Models, ed. L. Crivellari, I. Hubeny, D.G. Hummer, (Dordrecht: Kluwer) 440

Kurucz, R.L. 1992, in IAU Symp. 149, The Stellar Populations of Galaxies, ed. B. Barbuy, A. Renzini, (Dordrecht: Kluwer), 225

Lemoine, M., Ferlet, R., Vidal-Madjae, A., Emerich, C. & Bertin, P. 1993, A&A, 269, 469

Magazzu, A., Rebolo, R. and Pavlenko, Ya.V. 1992, ApJ, 392, 159

Mestel, L. 1953, MNRAS, 113, 716

Mestel, L. 1984, in $3^{rd}$ Cambridge Workshop on Cool Stars, Stellar Systems and the Sun, ed. S.L. Baliunas & L. Hartmann (New York: Springer) 49

Mestel, L. & Weiss, N.O. 1987, MNRAS, 226, 123

Michaud, G. 1986, ApJ, 302, 650

Michaud, G. and Charbonneau, P. 1991, Space.Sci.Rev., 57, 1

Michaud, G. & Proffitt, C.R. 1993, in Inside the Stars, IAU Col. 137, ed. A. Baglin & W.W. Weiss (San Fransico: ASP), 246

Pinsonneault, M.H. 1994, to appear in Proceedings of the 8th Cambridge Cool Stars Workshop, ed. J.P. Caillault

Pinsonneault, M.H., Kawaler, S.D., & Demarque, P. 1990, ApJS, 74, 501

Pinsonneault, M.H., Kawaler, S.D., Sofia, S., & Demarque, P. 1989, ApJ, 338, 424

Press, W.H. 1981, ApJ, 245, 286

Proffitt, C.R. & Michaud, G. 1991, ApJ, 371, 584

Radick, R.R., Thompson, D.T., Lockwood, G.W., Duncan, D.K. & Baggett, W.E. 1987, ApJ, 321, 459

Schatzman, E. 1962, Ann. d'Ap. 25, 18

Skumanich, A. 1972, ApJ, 171, 565

Soderblom, D.R. & Mayor, M. 1993, 105, 226

Soderblom, D.R., Fedele, S.B., Jones, B.F., Stauffer, J.R., & Prosser, C.F. 1993b, AJ, 106, 1080

Soderblom, D.R., Jones, B.F., Balachandran, S., Stauffer, J.R., Duncan, D.K., Fedele, S.B. & Hudon, J.D. 1993a, AJ, 106, 1059

Soderblom, D.R., Pilanchowski, C.A., Fedele, S.B. & Jones, B.F. 1993d, AJ, 105, 2299

Soderblom, D.R., Stauffer, J.R., Hudon, J.D., & Jones, B.F. 1993c, ApJS, 85, 315

Spite, M., Spite,. F., Peterson, R. C., & Chaffee, F. H. Jr. 1987, A&A (Letters), 172, L9

Spruit, H.C. 1987, in The Internal Solar Angular Velocity, ed. B. Durney & S. Sofia (Dordrecht: Reidel), 185

Stauffer, J.R. & Hartmann, L.W. 1987, ApJ, 378, 333

Sweet, P.A. 1950, MNRAS, 110, 548

Swenson, F.J. & Faulkner, J. 1992, ApJ, 395, 654

Swenson, F.J., Faulkner, J., Iglesias, C.A., Rogers, F.J. & Alexander, D.R. 1994, ApJ, 422, L79

Thorburn, J.A., Hobbs, L.M, Deliyannis, C.P. & Pinsonneault, M.H. 1993, ApJ, 415, 150

Tomkin, J. & Popper, D.M. 1986, AJ, 91, 1428

Vauclair, S. 1983, in Astrophysical Processes in Upper Main Sequence Stars, ed. B. Hauck and A. Maeder, (Geneva Obs.:Geneva),167.

von Zeipel, H. 1924, MNRAS, 84, 665

Wasiutynski, J. 1946, Ap. Norvegica, 4, 1

Zahn, J.-P. 1974, in IAU Symp. 59, Stellar Instability and Evolution, ed. P. Ledoux, A. Noels & A.W. Rodgers (Dordrecht:Reidel) 185

Zahn, J.-P. 1993, in Astrophysical Fluid Dynamics, Les Houches XLVII, ed. J.-P. Zahn & J. Zinn-Justin (New York: Elsevier Science Pub.), 561